\documentclass[11pt]{article}
\usepackage{amsmath}
\usepackage{amsfonts}
\usepackage{amssymb}

\newtheorem{theorem}{Theorem}[section]
\newtheorem{lemma}[theorem]{Lemma}

\newtheorem{definition}[theorem]{Definition}

\title{Uniqueness of Gibbs States of a Quantum System on Graphs}

\author{Dorota K\c{e}pa \\ Instytut Matematyki UMCS, 20-031 Lublin, Poland \\
DKEPA@golem.umcs.lublin.pl  \\[2ex]
         Yuri Kozitsky\thanks{ Supported by the DFG under the Project 436 POL
113/115/0-1}
                      \\ Instytut Matematyki UMCS, 20-031 Lublin, Poland \\
              jkozi@golem.umcs.lublin.pl  \\  }

\begin{document}

\maketitle

\begin{abstract}
Gibbs states of an infinite system of interacting quantum particles
are considered. Each particle moves on a compact Riemannian manifold
and is attached to a vertex of a graph (one particle per vertex).
Two kinds of graphs are studied: (a) a general graph with locally
finite degree; (b) a graph with globally bounded degree. In case
(a), the uniqueness of Gibbs states is shown under the condition
that the interaction potentials are uniformly bounded by a
sufficiently small constant. In case (b), the interaction potentials
are random. In this case, under a certain condition imposed on the
probability distribution of these potentials the almost sure
uniqueness of Gibbs states has been shown.

\end{abstract}

\section{Introduction}

Let $(M, \sigma)$ be a compact Riemannian manifold with the
corresponding normalized Riemannian volume measure $\sigma$. Let
also $\mathsf{G} = (\mathsf{L}, \mathsf{E})$ be a (countable) graph
with no loops, isolated vertices, and multiple edges. The model we
consider in this article  describes a system of interacting quantum
particles, each of which is attached to a vertex $\ell \in
\mathsf{L}$ (one particle per vertex), its position is described by
$q_\ell \in M$.The formal Hamiltonian of the model is
\begin{equation} \label{1}
H = - \frac{\hbar^2}{2m} \sum_{\ell \in \mathsf{L}} \mathit{\Delta}_\ell -
\sum_{\langle \ell, \ell' \rangle \in \mathsf{E}} v_{\ell\ell'} (q_\ell, q_{\ell'}),
\end{equation}
where $m$ is the particle mass, $\mathit{\Delta}_\ell$ is the
Laplace-Beltrami operator with respect to $q_\ell$, and $v_{\ell
\ell'}$ is a symmetric continuous function on $M\times M$. By
$\langle \ell , \ell' \rangle$ we denote the edge of the graph which
connects the named vertices.   The Hamiltonian of the free particle
\begin{equation}
\label{2}
H_\ell = - \frac{\hbar^2}{2m} \mathit{\Delta}_\ell
\end{equation}
is a self-adjoint operator in the physical Hilbert space
$\mathcal{H}_\ell = L^2 (M, \sigma)$. It is such that
\begin{equation*}
{\rm trace} \exp(- \tau H_\ell) < \infty
\end{equation*}
for all $\tau>0$. If one sets the periodic conditions at the
endpoints of the interval $[0,\beta]$, $\beta>0$ being the inverse
temperature, then the semi-group $\exp(- \tau H_\ell)$, $\tau \in
[0, \beta]$ defines a $\beta$-periodic Markov process, see
\cite{KL}, called a Brownian bridge, described by the Wiener bridge
measure. More on the properties of such operators and measures can
be found e.g. in \cite{Driver}.

A complete description of the equilibrium thermodynamic properties
of an infinite particle system may be made by constructing its Gibbs
states. In the quantum case, such states are defined as positive
normal functionals on the algebras of observables satisfying the
Kubo-Martin-Schwinger (KMS) conditions, which reflect the
consistency between the dynamic and thermodynamic properties of the
system proper to the equilibrium. Often, also in the case considered
here, the KMS conditions cannot be formulated explicitly, which
makes the problem of constructing the Gibbs states as KMS states
unsolvable. Since 1975, an approach based on the properties of the
semigroups generated by the local Hamiltonians is developed. In this
approach, in  \cite{AKMS1,AKMS2} the Gibbs states of the model
(\ref{1}), with $\mathsf{G}$ being a simple cubic lattice
$\mathbb{Z}^d$ with the edges connecting nearest neighbors, were
constructed in terms of probability measures on path spaces -- the
so called Euclidean Gibbs measures. So far, this is the only
possible way to define Gibbs states of such models and hence to give
a mathematical description of their thermodynamic properties. The
existence of Euclidean Gibbs measures in a simple manner follows
from the compactness of the manifold $M$. In the small mass limit,
which can also be treated as the strong quantum limit, the
uniqueness of Euclidean Gibbs measures at all $\beta$ was proven in
\cite{AKMS2}. The existence and uniqueness of the ground state,
i.e., of the Euclidean Gibbs measure at $\beta=+\infty$, was proven in
\cite{AKMS1}, also for small $m$. In both cases $\beta< +\infty$ and $\beta=+\infty$, 
the results were obtained 
by means of cluster expansions. The same method yields the proof of
the uniqueness for small $\beta$ (high temperature uniqueness), 
which is equivalent to the case of small 
\[
\sup_{\ell, \ell'} \sup_{q_\ell, q_{\ell'}} |v_{\ell, \ell'}(q_\ell,q_{\ell'})| 
\]
In the cluster expansion method, the properties of the lattice $\mathbb{Z}^d$
are crucial. Thus, it would be of interest to extend the latter result to
the case of more general graphs, especially those of globally unbounded degree.
Another possible extension would be considering random potentials.

In this article, we present two
statements establishing the uniqueness of Gibbs states of the model
(\ref{1}), which occurs due to weak interaction. In the first one
(Theorem \ref{1tm}), the graph $\mathsf{G}$ obeys a condition, by
which vertices of large degree should be at large distance of each
other. This is the only condition imposed on the graph -- we do not
suppose that it has globally bounded degree. In the second statement
-- Theorem \ref{2tm} -- we suppose such a boundedness, however the
functions $v_{\ell \ell'}$, $\langle \ell , \ell' \rangle \in
\mathsf{E} $, now are random. We claim that there exists a family of
the probability distributions on the space of these functions such
that, for every member of this family, the Gibbs state of the model
is almost surely unique. The proof of  these statements is based on
an extension of the method of \cite{BD}, where the single-spin
spaces were finite. Here we give its brief sketch. A detailed presentation of
this proof, as well as a more detailed explanation of the connection
between the Gibbs states and Euclidean Gibbs measures of the model
(\ref{1}), will be given in a separate publication.

\section{Setup and Results}

Since in our study the role of  the particle mass and the
temperature will be trivial, for notational convenience we set $m=
\beta= \hbar =1$. Throughout the paper, for a topological space
${Y}$, by $\mathcal{B}({Y})$ we denote the corresponding Borel
$\sigma$-field. By saying that $\mu$ is a measure on $Y$, we mean
that $\mu$ is a measure on the measure space $(Y, \mathcal{B}(Y))$.
For a measurable function $f:Y\rightarrow \mathbb{R}$, which is
$\mu$-integrable, we write
\[
\mu(f) = \int_Y f {\rm d}\mu.
\]

Given $\ell \in \mathsf{L}$,  we denote
\begin{equation}
\label{3} X_\ell = \{ x_\ell \in C([0, 1]\rightarrow M) \ | \ x_\ell
(0) = x_\ell (1)\},
\end{equation}
which is a metric space with the metric
\[ \varrho(x,y) = \sup_{\tau
\in [0,1]} d(x(\tau), y(\tau)),
\]
$d$ being the Riemannian  metric. This is our single-spin space. The
Wiener bridge measure $\chi_\ell$ is defined on $X_\ell$ by its
values on the cylinder sets
\begin{equation*}
B_{\tau_1, \dots, \tau_n} = \{ x_\ell \in X_\ell \ | \ x_\ell (\tau_1) \in B_1 , \dots x_{\ell} (\tau_n) \in B_n\},
\end{equation*}
where $B_i \in \mathcal{B}(M)$, $i=1, \dots , n$, and $\tau_i \in
[0,1]$ are such that $\tau_1 < \tau_2 \cdots < \tau_n$. For such a
set,
\begin{eqnarray}
\label{4} \chi_\ell (B_{\tau_1, \dots, \tau_n}) & = & \int_{B_1
\times \cdots \times B_n} p_{\tau_2 - \tau_1} (\xi_1,\xi_2) \times
\cdots \\ &\times & p_{\tau_n - \tau_{n-1}} (\xi_{n-1}, \xi_{n})
p_{1 + \tau_1 - \tau_{n}} (\xi_n ,\xi_1 )\sigma ({\rm d}\xi_1 )
\cdots \sigma({\rm d}\xi_n),   \nonumber
\end{eqnarray}
where $p_\tau (\xi, \eta)$, $\tau >0$ is the integral kernel of the
operator $\exp (-\tau H_\ell)$.

Let $\mathfrak{L}_{\rm fin}$ be the family of all finite non-void
subsets of $\mathsf{L}$. As usual, for $\Delta \subset \mathsf{L}$,
we use the notation $\Delta^c = \mathsf{L}\setminus \Delta$.  For
$\Delta \subseteq \mathsf{L}$,  the Cartesian product
\begin{equation}
\label{5}
X_\Delta = \prod_{\ell \in \Delta} X_\ell,
\end{equation}
is equipped with the product topology. Its elements are denoted by
$x_\Delta$;  we write $X = X_{\mathsf{L}}$ and $x = x_{\mathsf{L}}$.
Given $\langle \ell, \ell' \rangle \in \mathsf{E}$, let
$\mathit{\Omega}_{\ell \ell'}$ be a copy of the Banach space
$C(M\times M \rightarrow \mathbb{R})$ of continuous symmetric
functions equipped with the supremum norm. Then we set
\begin{equation}
\label{6}
\mathit{\Omega} = \prod_{ \langle \ell, \ell' \rangle \in \mathsf{E}} \mathit{\Omega}_{\ell\ell'},
\end{equation}
and equip this space with the product topology. Elements of
$\mathit{\Omega}$ are denoted by $v$. For $v_{\ell\ell'}\in
\mathit{\Omega}_{\ell\ell'}$ and given $x_\ell$, $x_{\ell'}$, we set
\begin{equation}
\label{7}
V_{\ell \ell'}(x_\ell, x_{\ell'}) = \int_0^1 v_{\ell \ell'}(x_\ell (\tau) , x_{\ell'}(\tau)) {\rm d}\tau.
\end{equation}
Then $V_{\ell \ell'}$ is a bounded continuous symmetric function on $X_\ell \times X_{\ell'}$. Obviously,
\begin{equation}
\label{8}
\|V_{\ell \ell'}\| \leq \|v_{\ell\ell'}\|,
\end{equation}
 where $\|\cdot \|$ stands for the supremum norm in both cases.

For $\Lambda \subset \mathsf{L}$, we set
\begin{equation}
\label{9}
\chi_\Lambda = \bigotimes_{\ell \in \Lambda} \chi_\ell,
\end{equation}
which is a probability measure on $X_\Lambda$. For such a $\Lambda$, we also set
\begin{eqnarray}
\label{10}
\mathsf{E}_\Lambda & =  &\{ \langle \ell, \ell' \rangle \in \mathsf{E} \ | \ \ell, \ell' \in \Lambda\},\\
\partial_{\mathsf{E}} \Lambda & = & \{ \langle \ell, \ell' \rangle \in \mathsf{E}
 \ | \  \ell \in \Lambda, \ \ell' \in \Lambda^c\},
\nonumber \\
\partial_{\mathsf{L}} \Lambda & = & \{\ell' \in \Lambda^c
 \ | \  \exists \ell \in \Lambda: \ \langle \ell, \ell' \rangle \in \mathsf{E}\}. \nonumber
\end{eqnarray}
The latter sets are called the edge and the vertex boundary of
$\Lambda$ respectively.

For $\Delta \subset \Delta'$, each element of $X_{\Delta'}$ can be
decomposed $x_{\Delta'} = x_\Delta \times x_{\Delta' \setminus
\Delta}$. In particular, one has $x = x_\Delta \times x_{\Delta^c}$.
Given $\Lambda \in \mathfrak{L}_{\rm fin}$, $y\in X$, and $B \in
\mathcal{B}(X)$, we set
\begin{equation}\label{11}
\pi_\Lambda (B|y) = \frac{1}{Z_\Lambda (y)} \int_{X_\Lambda}
 \mathbb{I}_B (x_\Lambda \times y_{\Lambda^c}) \exp\left[ 
 V_\Lambda (x_\Lambda|y)\right] \chi_\Lambda ({\rm d}x_\Lambda),
\end{equation}
where $\mathbb{I}_B$ is the indicator of $B$,
\begin{equation}
\label{12}
V_\Lambda (x_\Lambda|y) = \sum_{\langle \ell , \ell'\rangle \in 
\mathsf{E}_\Lambda} V_{\ell \ell'}(x_\ell , x_{\ell'} )
 + \sum_{\langle \ell,  \ell' \rangle \in \partial_{\mathsf{E}} \Lambda} V_{\ell \ell'} (x_\ell , y_{\ell.'}),
\end{equation}
and
\begin{equation}
\label{13}
Z_\Lambda (y) = \int_{X_\Lambda}  \exp\left[  V_\Lambda (x_\Lambda|y)\right] \chi_\Lambda ({\rm d}x_\Lambda).
\end{equation}
Thus, for every $v\in \mathit{\Omega}$, $\pi_\Lambda$ is a
probability kernel from $(X, \mathcal{B}(X))$ into intself. The
family $\{\pi_\Lambda\}_{\Lambda \in \mathfrak{L}_{\rm fin}}$ is the
local Gibbs specification of the model considered.
\begin{definition}\label{1df}
A probability measure $\mu$ on $X$ is called a Euclidean Gibbs
measure of the model (\ref{1}) if for every $\Lambda \in
\mathfrak{L}_{\rm fin}$ and any $B \in \mathcal{B}(X)$,
\begin{equation}
\label{14}
\mu(B) = \int_X \pi_\Lambda (B|x) \mu({\rm d}x).
\end{equation}\end{definition}
The set of all such Euclidean Gibbs measures will be denoted by
$\mathcal{G}$. If necessary, we write $\mathcal{G}(v)$ to indicate
the dependence on the choice of the interaction potentials.

For a topological space $Y$, by $\mathcal{P}(Y)$ we denote the set
of all probability measures on $Y$. We equip it with the usual weak
topology defined by all bounded continuous functions $f:Y\rightarrow
\mathbb{R}$, the set of which will be denoted by $C_{\rm b}(Y)$. One
can show that for every $f\in C_{\rm b}(X)$ and any $\Lambda\in
\mathfrak{L}_{\rm fin}$, $\pi_\Lambda (f|\cdot)\in C_{\rm b}(X)$;
thus, the local Gibbs specification has the Feller property. By the
latter property and by the compactness of $M$, one can show that for
every $x\in X$, the family $\{\pi_\Lambda(\cdot|x )\}_{\Lambda \in
\mathfrak{L}_{\rm fin}}\subset \mathcal{P}(X)$ is relatively compact
in the weak topology and that its accumulation points are elements
of $\mathcal{G}$. Hence, the latter set is non-void. For $v=0$, the
family $\{\pi_\Lambda\}_{\Lambda \in \mathfrak{L}_{\rm fin}}$ is
consistent in the Kolmogorov sense. Then by the Kolmogorov extension
theorem, $\mathcal{G}(0)$ is a singleton. We are going to prove that
this property persists if $v$ is nonzero but $\|v\|$ is small. Our
results cover the following two cases. Given $\ell \in \mathsf{L}$,
by $m_\ell$ we denote the degree of $\ell$, that is, $m_\ell = \#
\partial_{\mathsf{E}} \{\ell\}$. For $\ell , \ell' \in \mathsf{L}$,
by $\rho(\ell, \ell')$ we denote the distance between these vertices
-- the length of the shortest path connecting $\ell$ with $\ell'$.
In the first case, the only condition imposed on the graph
$\mathsf{G}$ is that for any $\ell, \ell'\in \mathsf{L}$,
\begin{equation}
\label{15}
\rho(\ell , \ell') \geq \phi \left(\min \{ m_\ell, m_{\ell'}\}\right),
\end{equation}
where $\phi:\mathbb{N}\rightarrow [1, +\infty)$ is a nondecreasing function, such that
\begin{equation}
\label{16}
\sum_{n=1}^{+\infty} \frac{n}{\phi(n)} < \infty.
\end{equation}
Set
\begin{equation}
\label{17}
\varkappa (v) = \sup_{\langle \ell , \ell' \rangle \in \mathsf{E}} 16 \left[\exp \left( 4 \|v_{\ell \ell'}\| \right) -1 \right],
\end{equation}
and, for $\delta >0$,
\begin{equation}
\label{18}
\mathit{\Omega}_\delta = \{ v \in \mathit{\Omega} \ | \ \varkappa (v) \leq \delta\}.
\end{equation}
\begin{theorem} \label{1tm}
There exists $\delta\in (0,1)$, depending on the choice of $\phi$, such that  for every $v\in \mathit{\Omega}_\delta$, the set $\mathcal{G}(v)$ is a singleton.
\end{theorem}
Our second result describes the case where the potential $v\in \mathit{\Omega}$ is random.
Let $\mathcal{Q}$ be the family of all probability measures on $\mathit{\Omega}$ having the product form
\begin{equation}
\label{19}
Q = \bigotimes_{\langle \ell , \ell' \rangle \in \mathsf{E}} Q_{\ell \ell'}, \quad Q_{\ell \ell'} \in \mathcal{P}(\mathit{\Omega}_{\ell\ell'}).
\end{equation}
Given $\lambda >0$ and $\theta \in (0,1)$, we set
\begin{equation}
\label{20}
\mathcal{Q}(\lambda ,\theta) = \{Q \in \mathcal{Q} \ | \ \forall \langle \ell , \ell' \rangle \in \mathsf{E}: \ \ Q_{\ell\ell'} (\| v_{\ell\ell'}\|>\lambda ) < \theta \}.
\end{equation}
\begin{theorem} \label{2tm}
Let the graph $\mathsf{G}$ be such that
\begin{equation} \label{21}
\sup_{\ell \in \mathsf{L}} m_\ell < \infty.
\end{equation}
Then there exist $\lambda >0$ and $\theta\in (0,1)$ such that, for every $Q\in \mathcal{Q}(\lambda , \theta)$, the set $\mathcal{G}(v)$ is a singleton for $Q$-almost all $v$.
\end{theorem}

\section{Comments on the Proof}

To prove the above results we will use the method developed in \cite{BD}, 
where similar statements were proven for a model of classical spins with finite single-spin spaces.
Thus, the only generalization we need is to extend this method to 
the case of the single-spin space $X_\ell$ as introduced above.

Given $\Delta \in \mathfrak{L}_{\rm fin}$, we set
\begin{equation}
\label{22} \nu_\Delta ({\rm d} x_\Delta) = \frac{1}{Z_\Delta}
\exp\left[ \sum_{\langle \ell , \ell' \rangle \in \mathsf{E}_\Delta}
V_{\ell \ell'} (x_\ell. x_{\ell'}) \right] \chi_\Delta ({\rm
d}x_\Lambda),
\end{equation}
where $1/Z_\Delta$ is a normalization factor. Thus,
 $\nu_\Delta$ is a probability measure on $X_\Delta$ -- the local Euclidean Gibbs
 measure. For $\Lambda \subset \Delta \subset \mathsf{L}$, we set
\begin{equation} \label{23}
\mathsf{E}_\Delta (\Lambda) = \partial_{\mathsf{E}} \Lambda \cap
\mathsf{E}_\Delta,
\end{equation}
i.e., $\mathsf{E}_\Delta (\Lambda)$ is the smallest subset of
$\mathsf{E}_\Delta$ along which one can cut out $\Lambda$ from
$\Delta$. Then, for $\Lambda \subset \Delta \in \mathfrak{L}_{\rm
fin}$, we have
\begin{equation}
\label{23a} \nu_\Delta ({\rm d}x_\Delta) = \nu_{\Delta} ({\rm
d}x_{\Delta \setminus \Lambda} |x_{ \Lambda}) \nu^\Lambda_{\Delta}
({\rm d}x_\Lambda),
\end{equation}
where $\nu^\Lambda_{\Delta}$ is the projection of $\nu_\Delta$ onto
$\mathcal{B}(X_\Lambda)$ and
\begin{eqnarray} \label{23b}
\nu_{\Delta} ({\rm d}x_{\Delta \setminus \Lambda} |x_{ \Lambda})& =
& \frac{1}{N_{\Lambda} (x_\Lambda)} \exp\left[\sum_{\langle  \ell,
\ell'\rangle\in \mathsf{E}_{\Delta \setminus \Lambda} \cup
\mathsf{E}_\Delta (\Lambda)} V_{\ell \ell'}(x_\ell , x_{\ell'})
\right] \\ & \times & \chi_{\Delta
\setminus \Lambda} ({\rm d} x_{\Delta \setminus \Lambda}), \nonumber  \\
N_\Lambda (x_\Lambda) & = & \int_{X_{\Delta \setminus
\Lambda}}\exp\left[\sum_{\langle  \ell, \ell'\rangle\in
\mathsf{E}_{\Delta \setminus \Lambda} \cup \mathsf{E}_\Delta
(\Lambda)} V_{\ell \ell'}(x_\ell , x_{\ell'}) \right] \chi_{\Delta
\setminus \Lambda} ({\rm d} x_{\Delta \setminus \Lambda}).\nonumber
\end{eqnarray}
Given $\ell , \ell' \in \mathsf{L}$, by $\vartheta (\ell, \ell')$ we
denote a path, which connects these vertices. That is, $\vartheta
(\ell, \ell')$ is the sequence of vertices $\ell_0, \dots, \ell_n$,
$n \in \mathbb{N}$, such that: (a) $\ell_0 = \ell$ and $\ell_n =
\ell'$; (b) for all $j= 0, \dots, n-1$, $\langle \ell_{j+1}, \ell_j
\rangle \in \mathsf{E}$. The path $\vartheta(\ell , \ell')$ will be
called admissible if: (c) $m_{\ell_k} \geq 2$ for all $k = 1, \dots
, n-1$; (d) if $\ell_i = \ell_j$ for certain $i<j$, then there
exists a $k$, $i< k< j$, such that $m_{\ell_k} \geq m_{\ell_i} =
m_{\ell_j}$. The length of a path $|\vartheta (\ell, \ell')|$ is the
number of vertices in it, i.e., $|\vartheta (\ell, \ell')|=n$.

For $\langle \ell , \ell' \rangle \in \mathsf{E}$, we set, c.f.,
(\ref{17}),
\begin{equation} \label{24}
\varkappa_{\ell \ell'} = 16 \left[\exp \left( 4 \|v_{\ell
\ell'}\|\right) - 1  \right].
\end{equation}
 For a path $\vartheta(\ell  , \ell')$, we define
\begin{equation} \label{25}
R[\vartheta(\ell  , \ell')] = 4^{\varsigma(\ell) + \varsigma
(\ell')} \prod_{i=0}^{n-1} \varkappa_{\ell \ell'},
\end{equation}
where $\varsigma(\ell) = -1$ if $m_\ell =1$ and $\varsigma(\ell) =
0$ if $m_\ell \geq 2$. Finally, for $\Delta \subset \mathsf{L}$, we
set
\begin{equation} \label{26}
S_\Delta (\ell , \ell') = \sum R[\vartheta(\ell  , \ell')],
\end{equation}
where the summation is performed over all admissible paths
connecting $\ell$ with $\ell'$. The main element of the proof of
both our theorems is the following, lemma which is an adaptation of
Theorem 1 of \cite{BD} to the model considered here.
\begin{lemma} \label{1lm}
Given $\Lambda \in \mathfrak{L}_{\rm fin}$, let $f:X_\Lambda
\rightarrow \mathbb{R}$ be a bounded continuous function. Then for
every $\Delta \in \mathfrak{L}_{\rm fin}$, such that $\Lambda
\subset \Delta$, and arbitrary $\ell' \in \Delta \setminus \Lambda$,
$x_{\ell'}, x'_{\ell'} \in X_{\ell'}$, it follows that
\begin{equation} \label{27}
\left\vert \frac{\nu_\Delta (f|x_{\ell'})}{\nu_\Delta
(f|x'_{\ell'})} - 1 \right\vert \leq \sum_{\ell \in \Lambda}
S_\Delta (\ell , \ell').
\end{equation}
\end{lemma}
{\em Scetch of the proof:} The proof of the lemma is based on the
following estimates. Let $(Y, \mathcal{B}(Y), P)$ be a probability
space and $a, b, c$ be positive measurable real valued functions on
$Y$. Then
\begin{equation} \label{28}
\inf_{y\in Y}\left( \frac{b(y)}{c(y)}\right) \leq \frac{\int a(y)
b(y) P({\rm d}y)}{\int a(y) c(y) P({\rm d}y)}\leq \sup_{y\in
Y}\left( \frac{b(y)}{c(y)}\right),
\end{equation}
\begin{equation} \label{29}
\left\vert \frac{\int a(y) b(y) P({\rm d}y)}{\int a(y) c(y) P({\rm
d}y)} - 1\right\vert \leq \sup_{y\in Y} \left\vert \left(
\frac{b(y)}{c(y)}\right) -1\right\vert.
\end{equation}
Now let $a,b,c$ be as above, let also $a': Y \rightarrow (0+\infty)$
and this $a$ obey the conditions $P(a) = P(a') =1$. Set
\[
S(b,c) = \max\{ \sup_{y\in Y} b(y); \sup_{y\in Y} c(y) \}, \quad \
I(b,c) = \min\{ \inf_{y\in Y} b(y); \inf_{y\in Y} c(y) \}
\]
and suppose that $b$ and $c$ are such that $I(b,c)>0$. Then
\begin{equation} \label{30}
\left\vert \frac{\int a(y) b(y) P({\rm d}y)}{\int a'(y) c(y) P({\rm d}y)} - 1 \right\vert \leq
\frac{S(b,c)}{I(b,c)} - 1.
\end{equation}
Suppose now that $c$ is such that there exists $y_0\in Y$ for which
$c(y_0) = \inf_{y\in Y}c(y)$. Let positive $\gamma$, $\epsilon$,
$\delta$ be such that
\[
\left\vert \frac{a(y)}{a'(y)} - 1 \right\vert \leq \gamma , \quad
\left\vert \frac{c(y)}{c(y_0)} - 1 \right\vert \leq \epsilon, \quad
\left\vert \frac{b(y)}{c(y)} - 1 \right\vert \leq \delta.
\]
Then
\begin{equation} \label{31}
\frac{\int a(y) b(y) P({\rm d}y)}{\int a'(y) c(y) P({\rm d}y)} \leq
\delta + \epsilon \gamma + \delta \epsilon \gamma.
\end{equation}
By means of the estimates (\ref{28}) -- (\ref{31}) the proof of the
lemma follows by the same inductive scheme which was used in the
proof of Theorem 1 in \cite{BD}. $\square$

The proof of both theorems stated above follows from Lemma \ref{1lm}
by means of similar arguments the proof of Theorems 2, 3, and 4 in
\cite {BD} was based on.

\end{document}